\documentclass[aps,prl,twocolumn,amsmath]{revtex4-1} %

\usepackage{ulem}

\usepackage{color,graphicx}
\usepackage{dcolumn}%
\usepackage{bm}%

\usepackage{amsmath}
\usepackage{amsfonts}
\usepackage{amssymb}
\usepackage{rotate}

\usepackage{subfigure}
\usepackage[latin1]{inputenc}

\renewcommand{\-}{\,-\,}

\let\oldmarginpar\marginpar
\renewcommand\marginpar[1]{\-\oldmarginpar[\raggedleft\tiny #1]%
{\raggedright\tiny #1}}

\begin{document}

\title{Correlated phases of bosons in the flat lowest band of the dice lattice}

\author{G. M\"oller}
\affiliation{TCM Group, Cavendish Laboratory, J.~J.~Thomson Avenue, Cambridge CB3 0HE, UK}

\author{N. R. Cooper}
\affiliation{TCM Group, Cavendish Laboratory, J.~J.~Thomson Avenue, Cambridge CB3 0HE, UK}

\date{\today}

\begin{abstract}
We study correlated phases occurring in the flat lowest band of the dice lattice model at flux density one half. We discuss how to realize the
dice lattice model, also referred to as the $\mathcal{T}_3$ lattice, in cold atomic gases. We construct the projection of the model to the lowest
dice band, which yields a Hubbard-Hamiltonian with interaction-assisted hopping processes. We solve this model for bosons in two limits. In the
limit of large density, we use Gross-Pitaevskii mean-field theory to reveal time-reversal symmetry breaking vortex lattice phases. At low density,
we use exact diagonalization to identify three stable phases at fractional filling factors $\nu$ of the lowest band, including a 
classical crystal at $\nu=1/3$, a supersolid state at $\nu=1/2$ and a Mott insulator at $\nu=1$.
\end{abstract}

\maketitle

The dice lattice \cite{Horiguchi:1974p2543, Sutherland:1986p2542}, which is also referred to as the $\mathcal{T}_3$ lattice \cite{Vidal:1998p72}, 
gives rise to an exceptional mechanism for localization: when subjected to a magnetic field with a flux density $n_\phi$
of one half flux quantum per plaquette, it realizes ``Aharonov-Bohm cages'' that perfectly confine particle motion by destructive
interference around individual plaquettes \cite{Vidal:1998p72}. The presence of these localized states translates into a macroscopic 
degeneracy of states or equivalently a spectrum with flat energy bands.
A natural implementation of this model can be found in Josephson junction arrays \cite{Abilio:1999p1877}, where condensates of Cooper-pairs
on each site are well characterized by an $xy$-model for the classical order parameter. The rich physics of these systems stems from a highly degenerate manifold of states at
low energies \cite{Korshunov:2001p60}, described as vortex lattices. Dynamics in this manifold is slow \cite{Cataudella:2003p81},
and ordering at the lowest temperatures is determined by such subtle effects
as magnetic interactions of currents \cite{Korshunov:2004p57} 
or anharmonic fluctuations \cite{Korshunov:2005p1876}.

Given the perfectly flat bands in the spectrum of the dice lattice, interactions can potentially lead to strongly correlated
states in the regime of low particle density where number fluctuations are significant. This is reminiscent of the physics of the fractional
quantum Hall effect \cite{PrangeGirvin}, which is predicted to exist also in systems of bosonic atoms in the continuum \cite{cw}, or in
the Hofstadter bands of a square lattice \cite{Sorensen:2005p58,Palmer:2006p63,Moller:2009p184}. 
Similarly for fermions, there has been intense interest in realizing fractional quantum Hall states in general flat-band models with nonzero Chern numbers
\cite{Tang:2011p1780,*Sun:2011p1781,*Neupert:2011p1803,*Roy:2011p1779,*Regnault:2011p2571}.
However, the dice lattice model at flux density of one half is fundamentally different from this Landau level physics, 
as it does not break time-reversal symmetry. In this model, the role of interactions has been described only for the two-body problem of fermions, 
where it leads to delocalized zero-energy states of pairs \cite{Vidal:2000p1875,Vidal:2001p1874}.

In this paper, we study the physics of strongly correlated states in the flat lowest band of the dice lattice model at flux density $n_\phi =1/2$.
For additional motivation of this study, we sketch candidates for robust implementations of the dice lattice model in cold atoms.
In constrast to other flat band models, as on the kagomé lattice \cite{Huber:2010p814}, our proposal benefits from the magic of Aharonov-Bohm
cages and thus yields a perfectly flat lowest band separated by a large gap from higher bands, using only nearest-neighbor hopping. %
To analyse the problem of interacting particles in this flat band, we introduce an effective model obtained by projecting the Hamiltonian to this band, which is characterized by one dimensionless interaction parameter. In the limit of large density, we solve this model using Gross-Pitaevskii mean-field theory and find time-reversal breaking ground states 
that realize the same phase patterns as in the ground states of the classical $xy$-model \cite{Korshunov:2001p60}, but which have additional density modulations.
The projected model also enables efficient numerical studies of the many body physics on the dice lattice based on exact diagonalization. 
We undertake a numerical study in the regime of low particle density that is of interest for cold atomic gases on lattices. We identify stable phases of the model 
as a function of density and find evidence for several phases including classical crystalline states, a supersolid state with triangular crystalline order and a Mott insulator.

%
%
%
%

%

\begin{figure}[ttt]
\begin{center}
\includegraphics[width=0.94\columnwidth]{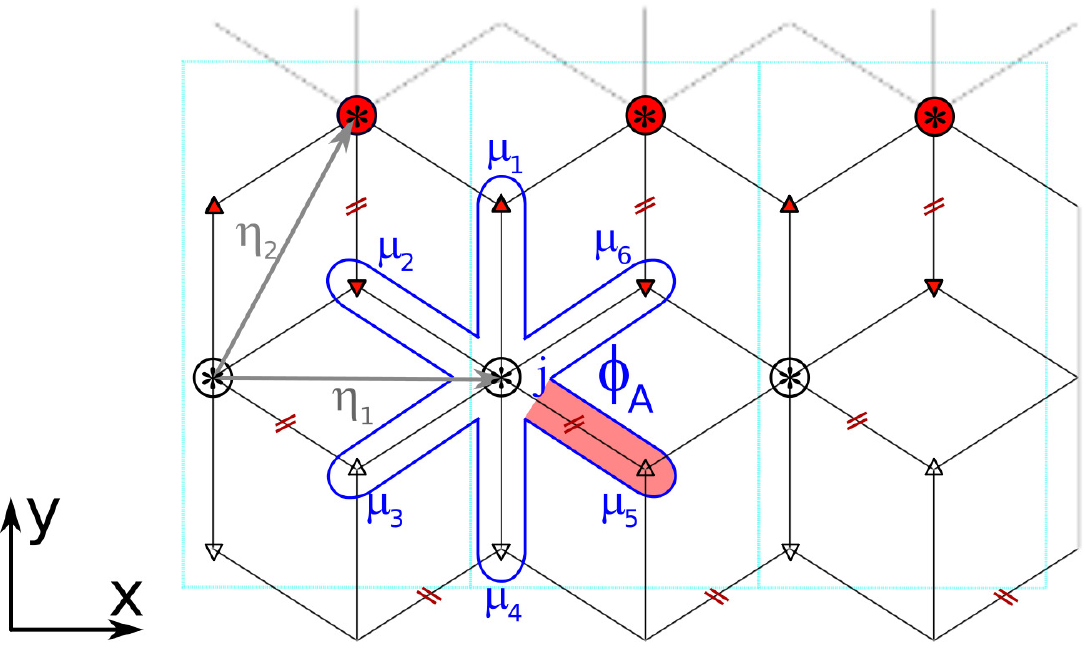}
\caption{(Color online) Figure showing the structure of the dice lattice, highlighting a rectangular unit cell and negative hoppings on three bonds (double hatched) that
realize the case of flux density $n_\phi=\frac{1}{2}$. The blue contour highlights the extent of a maximally localized single-particle wavefunction in the flat lowest band. 
Localized states centered around all six-fold connected sites of the lattice (encircled) form an orthonormal basis for this band, and lie on a triangular lattice. 
}
\label{fig:Structure}
\end{center}
\end{figure}

%
%
An optical dice lattice can formed by three mutually phase coherent pairs of counterpropagating laser beams at relative angles of $2\pi/3$ \cite{[{See e.g.,~}] Rizzi:2006p65,*Burkov:2006p70,*Bercioux:2009p441}. 
Here, we propose instead to use a set of three mutually incoherent pairs in the same geometrical arrangement.
This enables a particularly convenient scheme to be realized
for Yb by using an ``antimagic'' wavelength, such that two internal states ($^1S_0$ and $^3P_0$) are trapped at the points of maximum or minimum laser intensity \cite{Gerbier:2010p630,*GerbierPrivate}: these points again form
a dice lattice, with the structure shown in Fig.~\ref{fig:Structure}. %
To %
induce dynamics in this lattice, one employs laser-assisted hopping %
that simultaneously imprints phases onto the hopping matrix elements \cite{Gerbier:2010p630}.
For the flux density, $n_\phi=1/2$, that yields the desired flat band structure on the dice lattice \cite{Vidal:1998p72}, the magnetic unit cell contains six inequivalent atoms. 
Following the ideas of \cite{Gerbier:2010p630}, the required phases can be imprinted by optically assisted tunnelling involving several %
coupling lasers and one additional superlattice laser to break inversion symmetry of the magnetic unit cell.
We note that at the field strength $n_\phi=1/2$ time-reversal symmetry is not broken, so it is possible to choose a gauge with real tunnelling amplitudes and with only three bonds in the unit cell having a negative sign, as indicated in Fig.~\ref{fig:Structure}.  
This allows for an alternative potential implementation, 
using an optical dice lattice for single species atoms \cite{Burkov:2006p70}.
Negative hopping can be achieved in principle by shaking lattice sites \cite{Eckardt:2005p2066,*Lignier:2007p2067,*Eckardt:2010p2065}; the application in this case may be challenging, as it requires the three-fold connected sites in the magnetic unit cell to be independently shaken. Finally, the synthetic gauge field can also be simulated by the Coriolis force due to a rapid rotation of the system \cite{Tung:2006p2061,*Williams:2010p394,[{}] [{, for a review.}] Cooper:2008p250}.

Our main focus concerns the many-body physics of bosons on the dice lattice at this flux density, $n_\phi=1/2$,
as described by the Bose-Hubbard model %
\begin{align}
\label{eq:HubbardHam}
\mathcal{H} = & - t \sum_{<j,\mu>} \left( \hat a^\dagger_\mu \hat a_j e^{iA_{\mu j}} + h.c. \right) \\ 
& + \frac{U_\vartriangle}{2} \sum_{\mu\in \vartriangle,\triangledown} \hat a^\dagger_\mu \hat a_\mu(\hat a^\dagger_\mu \hat a_\mu-1) + \frac{U_\ast}{2} \sum_{j \in \ast} \hat a^\dagger_j \hat a_j(\hat a^\dagger_j \hat a_j-1).\nonumber
\end{align}
Here, $\hat a^{(\dagger)}$ are the creation/annihilation operators of atoms on single lattice sites. 
We adopt Roman indices for six-fold connected `hubs' ($\ast$) and use Greek indices to number threefold ($\vartriangle,\triangledown$) connected sites of the dice 
lattice. These types of site can have distinct on-site repulsion $U_\ast$, $U_\vartriangle$.
We choose a real gauge with $A_{\mu j}$ %
$= 0$ or $\pi$, with negative bonds highlighted in Fig.~\ref{fig:Structure}.

\begin{figure}[ttt]
\begin{center}
\includegraphics[width=0.93\columnwidth]{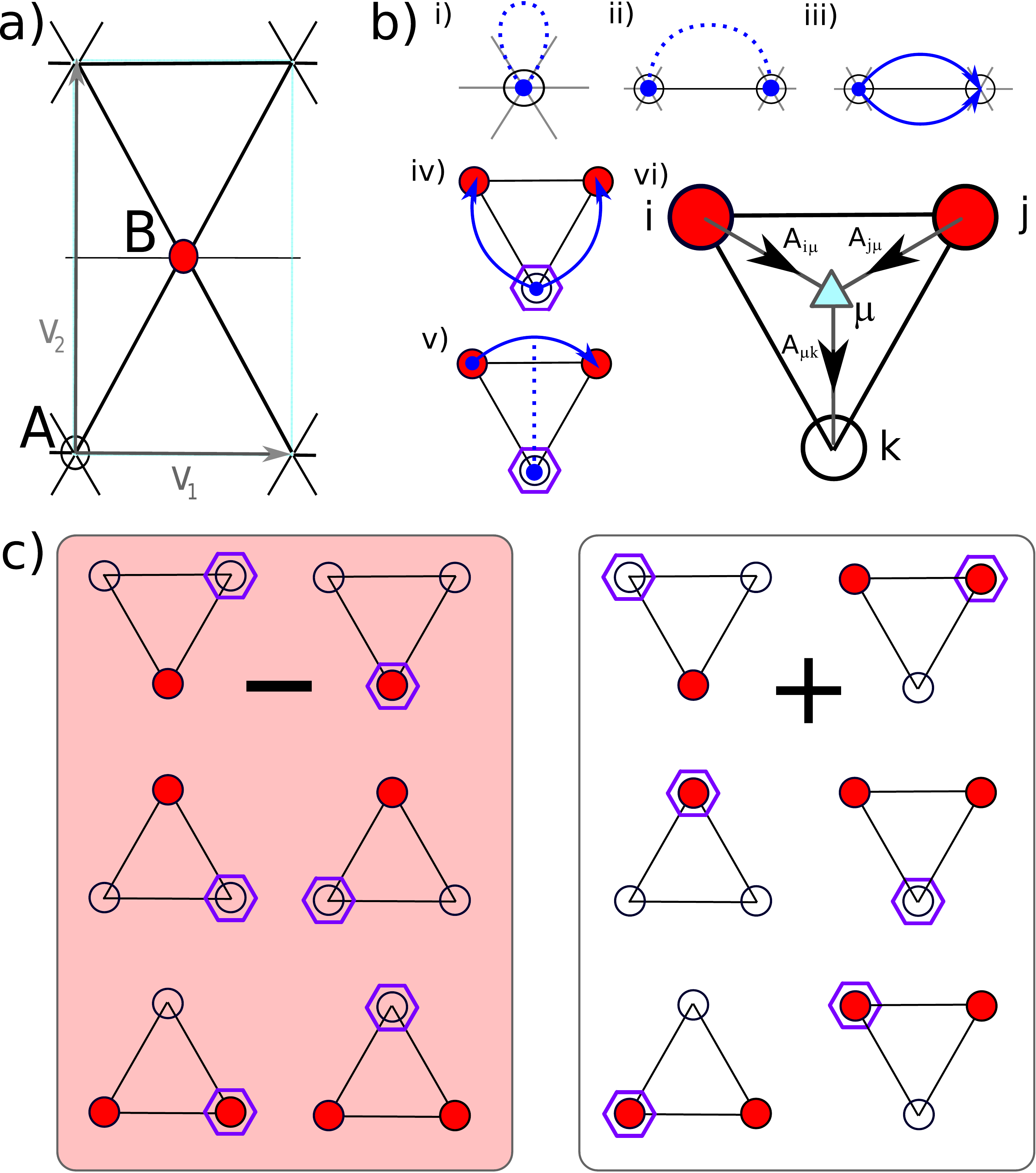}
\caption{(Color online) (a) The projective dynamics within the low energy band of the dice lattice at flux $n_\phi=1/2$ realizes an effective model on a triangular lattice with an enlarged magnetic unit cell of two distinct triangular sites A and B. (b) The density-density interactions of the dice Hamiltonian (\ref{eq:HubbardHam}) give rise to the five distinct processes i)-v), including an on-site (i) and nearest-neighbor (ii) interactions, coherent hopping of pairs of particles onto the same (iii) and two distinct neighboring sites (iv), and a stimulated hopping process (v). Hoppings are shown as arrows, density interactions as dotted lines. Processes (iv) and (v) have a nontrivial dependency on the gauge from (vi), see main text. c) Signs for the processes (iv,v) in the real gauge of Fig.~\ref{fig:Structure}: these processes involve sites on a triangular plaquette, of which there are four types classified by the participating sublattice indices [empty (A) / full (B) circles]; in addition one of the three sites carries two creation/annihilation operators (marked by a hexagon).}
\label{fig:EffectiveModel}
\end{center}
\end{figure}

The single-particle spectrum of (\ref{eq:HubbardHam}) at $n_\phi=1/2$ is given by three flat bands with $E=-\sqrt{6}t,0,\sqrt{6}t$ \cite{Vidal:1998p72}. 
We focus on the regime with $nU_{\ast/\vartriangle}/t \ll 1$, such that all dynamics occurs in the lowest band only. 
Wavefunctions localized on the hubs and extending to the neighboring threefold connected sites (see Fig.~\ref{fig:Structure}) span an orthonormal 
basis $\{\phi_j\}$ for the lowest band \cite{Vidal:2001p1874,NoteLocalization}. %
These $\phi_j$ have amplitude $1/\sqrt{2}$ on the central site $j$, and $1/\sqrt{12}\exp[i A_{j\mu}]$ on the peripheral sites $\mu$ \footnote{The projection to the upper band is similar, as single-particle states differ by the sign of the central site $j$ only.}.
In our real gauge, $\phi_j$ are identical up to translation in all unit cells.
(Both sublattices are also related by a magnetic translation involving a gauge transformation.) 
The projection onto the lowest band realizes a new effective problem on a triangular lattice visualized in Fig.~\ref{fig:EffectiveModel}, which derives from the density-density
interactions of the microscopic Hamiltonian (\ref{eq:HubbardHam}) via $V_{ijkl}= U_\vartriangle \sum_{\mu} \phi^*_i(\mu) \phi^*_j(\mu) \phi_k(\mu) \phi_l(\mu) + U_\ast \sum_{q} \phi^*_i(q) \phi^*_j(q) \phi_k(q) \phi_l(q)$,
and that reads 
\begin{align}
\label{eq:ProjectedH}
\mathcal{H}_\text{proj} = & \gamma_1 \sum_i \hat n_i(\hat n_i-1) + \gamma_2 \sum_{\langle i,j \rangle}\left[ \hat n_i \hat n_j 
+ \hat c^{\dagger 2}_i \hat c_j^2  +  \hat c^{\dagger 2}_j \hat c_i^2 \right] \nonumber\\
  + & \gamma_3 \sum_{\triangle(i,j,k)} \left[ \sigma^{ij}_{kk} \hat c^\dagger_i \hat c^\dagger_j \hat c_k^2 + \sigma^{ik}_{jk} \hat c^\dagger_i \hat c_j \hat n_k + h.c. \right],
\end{align}
with $\hat c^{\dagger}_j$ creating an atom in orbital $\phi_j$, and $\hat n_j = \hat c^{\dagger}_j\hat c_j$, $\gamma_1 = \frac{1}{8}U_\ast + \frac{1}{48}U_\vartriangle$, $\gamma_2 = \frac{1}{144} U_\vartriangle$, $\gamma_3 = \frac{1}{288}U_\vartriangle$. The prefactors 
$\sigma^{qr}_{st} = \exp i[ A_{q\mu} + A_{r\mu} + A_{\mu s} + A_{\mu t}]$ derive from the gauge fields around a shared threefold connected site $\mu$ [see Fig.~\ref{fig:EffectiveModel}(b-vi)].  
In the real gauge, $\sigma^{ij}_{kk}=\sigma^{ik}_{jk}$ reduces to a sign as illustrated in Fig.~\ref{fig:EffectiveModel}(c). This effective model includes on-site and nearest-neighbor interactions as well as pair hoppings, and interaction-assisted hopping processes. Equivalent projected models can be written generally also for multicomponent bosons/fermions.

Importantly, note that while the original model (\ref{eq:HubbardHam}) is strictly local, the projected Hamiltonian involves nonzero range 
interactions between nearest-neighbor sites. In contrast to the flat-band physics of the kagomé lattice \cite{Huber:2010p814}, longer range interactions vanish exactly.
The ratio of on-site terms to nearest-neighbor terms can be tuned, since $u=U_\ast/U_\vartriangle$ remains as a free parameter of the model (while the tunnelling $t$ drops 
out of the problem). Choosing $u\gg1$ while maintaining $ nU_{\ast,\vartriangle} \ll t$ defines the hardcore limit of the projected model, eliminating 
configurations involving double occupancy. 
We now inquire into the nature of the many-body ground states of spinless bosons as a function of density.
At densities $n\gg 1$, 
but remaining in the regime $ nU_{\ast,\vartriangle} \ll t$ governed by %
the projected Hamiltonian (\ref{eq:ProjectedH}), we expect that correlations 
can be neglected. We analyze this regime using a Gross-Pitaevskii mean-field equation deriving from the matrix elements $V_{ijkl}$ in the lowest band. This ansatz is based on
a condensate wavefunction which is a
coherent state of the form $|\Psi\rangle = \exp[\sum_j \alpha_j \hat c_j^\dagger] |0\rangle $, such that  $\hat c_j |\Psi\rangle = \alpha_j |\Psi\rangle$. 
Introducing a chemical potential $\mu$, %
the energy is
$\langle H \rangle = \frac{1}{2}\sum_{ijkl} V_{ijkl} \alpha^*_i \alpha^*_j \alpha_k \alpha_l - \mu \sum_j |\alpha_j|^2$.
Using steepest descent numerical minimization of the energy, and expanding the resulting states 
on the full dice lattice,
we find that the phase-patterns minimizing the energy are identical to those described by Korshunov for the $xy$-model \cite{Korshunov:2001p60}. These ground states break time-reversal symmetry and feature clusters of three plaquettes sharing the same vorticity or phase winding $\pm \pi$. Unlike for the $xy$-model, the density for these mean-field states of the projected model (\ref{eq:ProjectedH}) is not homogeneous. Instead, we find $n_\ast = 2 n_\vartriangle$ for $u=1$. %
Quantum fluctuations are likely to break the degeneracy of these different vortex patterns, as occurs for thermal fluctuations in Josephson junction arrays \cite{Korshunov:2005p1876}. However, this effect is beyond Gross-Pitaevskii theory, which does not capture the role of quantum fluctuations.

\begin{figure}[tbhp]
\begin{center}
\includegraphics[width=0.94\columnwidth]{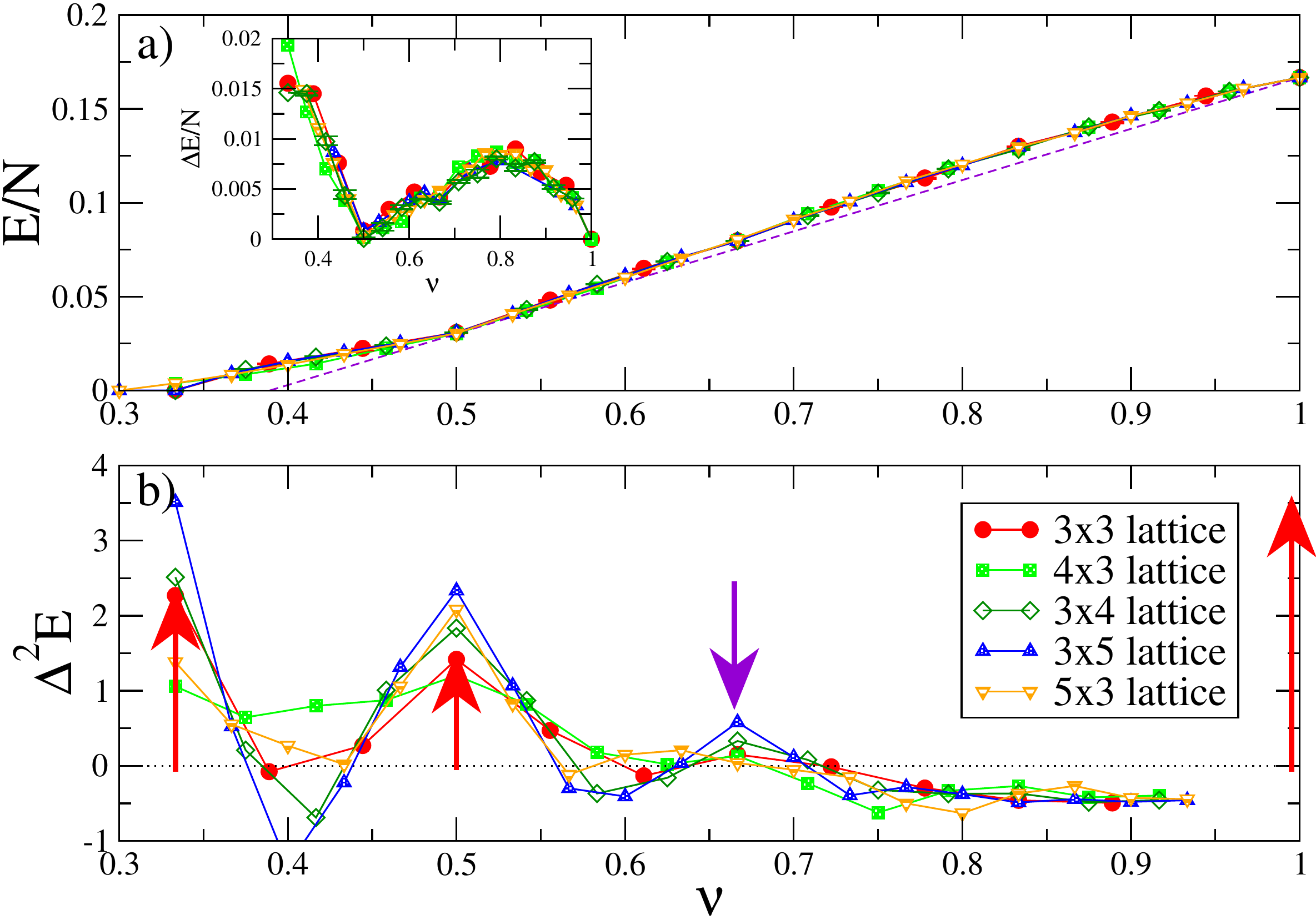}
\caption{(Color online) (a) Energy per particle as a function of the filling factor $\nu=3n$ of the low-lying band for the projected Hamiltonian (\ref{eq:ProjectedH}) 
at $n_\phi=1/2$ with hardcore interactions. Finite size effects between different lattice geometries and the effect of twisted boundary conditions (band-width shown as error bars for $3\times 3$ and $3\times 4$ geometries) are small. Inset: Energy measured with respect to the tangent at points $\nu=1/2$ and $\nu=1$ 
(dashed line in main panel), showing that $\nu=1/2$ is the only stable phase between the crystal at $\nu=1/3$ and Mott state at $\nu=1$ for $u\gg 1$. (b) Second difference of the energy per particle. Maxima indicate potential incompressible phases, including the states at $\nu=1/3$, $1/2$ and $1$ as well as an additional weaker candidate at $\nu=2/3$ that is overridden by phase separation.
}
\label{fig:GSEnergy}
\end{center}
\end{figure}

To address the question of how fluctuations resolve the frustration seen in the mean-field solutions, we study the regime of low particle density that takes the system into the strongly correlated limit. %
We express $n$ in terms of the density of states in the lowest band $n_g$, defining the band filling factor $\nu=n/n_g=3n$.
A remarkable feature of the Hamiltonian (\ref{fig:EffectiveModel}) is that all hoppings are mediated only by the presence of neighboring particles, while single
particles remain stationary. Consequently, the many-body spectrum at low filling factor $\nu<\nu_c=1/3$ has a highly degenerate zero-energy ground state. 
The two-body problem has been studied for two-species fermions with repulsive contact interactions \cite{Vidal:2001p1874}, 
and it features delocalized zero-energy spin-singlet states in addition to the trivial zero-energy states with particles placed at a distance of more than one lattice vector. The two-boson wavefunctions have the same structure, due to their identical spatial symmetry. Counting $E=0$ states in the spectra of finite size systems confirms this.
For filling factors $\nu<\nu_c=1/3$ the system has infinite compressibility.
Precisely at $\nu_c$, the model yields an incompressible ground state that %
is a classical crystal of triangular symmetry with unit vectors $\vec u_1=\vec \eta_1 + \vec \eta_2$ and $\vec u_2=2\vec \eta_1 - \vec \eta_2$ 
(see Fig.~\ref{fig:Structure} for the definition of $\vec\eta_{1,2}$). The three degenerate crystal ground states can be written as 
$|\Psi_c\rangle = \prod_{n,m} \hat c^\dagger[\vec r_t + n \vec u_1 + m \vec u_2] |0\rangle$, 
where $\hat c^\dagger[\vec r]$ fills the orbital centered at %
$\vec r$, and $\vec r_t \in \{\vec 0, \vec \eta_1, \vec \eta_2\}$ are translations with respect to the origin. Compression 
to densities $\nu>\nu_c=1/3$ costs a nonzero energy. Such incompressible states at noninteger filling can exist only due to the nonlocal interactions of the effective model.

\begin{figure*}[tbhp]
\begin{center}
\includegraphics[width=1.7\columnwidth]{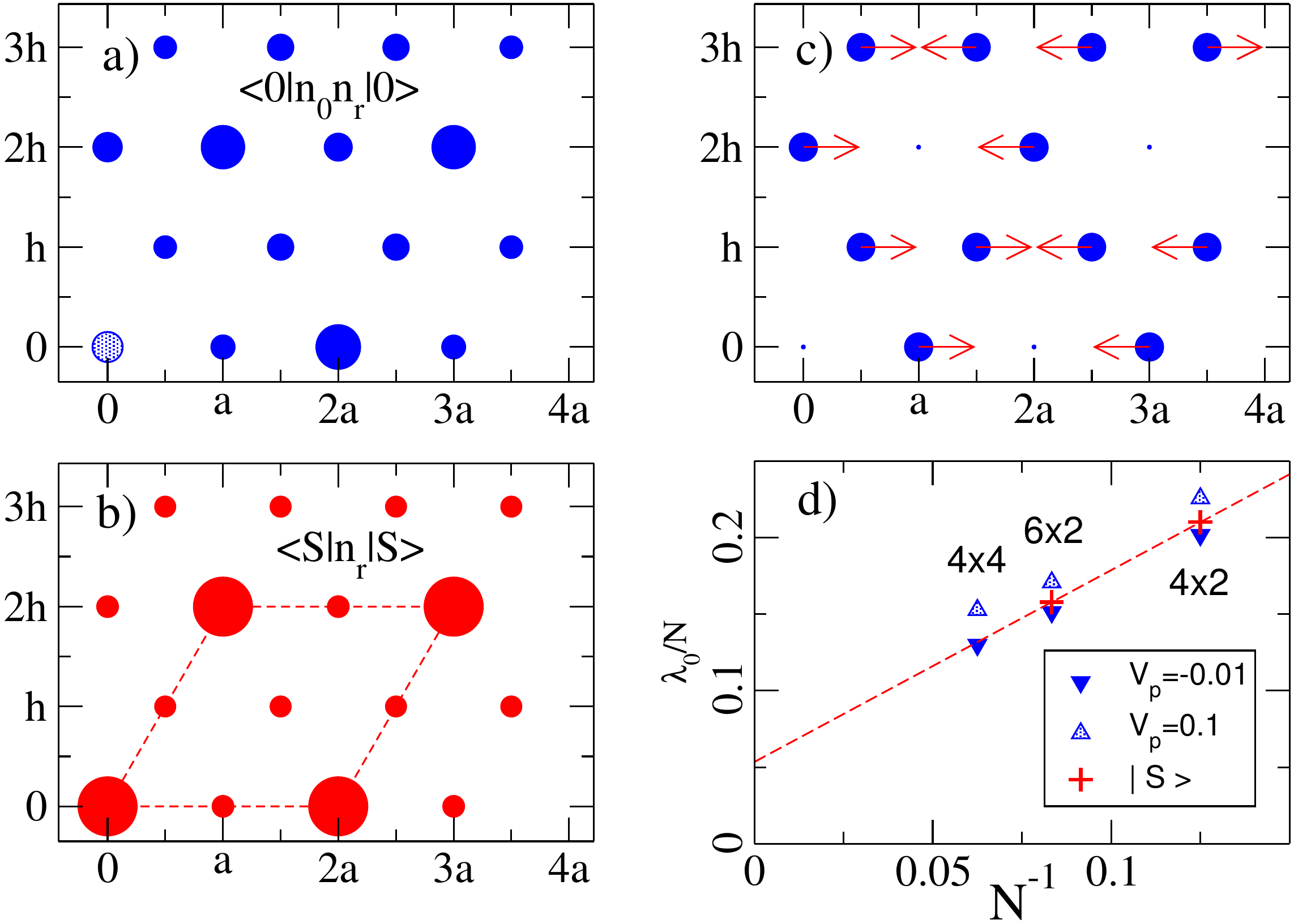}
\caption{(Color online) Data for $N=8$ atoms at $\nu=1/2$ on a lattice of $4\times 2 $ unit cells. (a) Two-point correlation function $\langle\hat n_r \hat n_0 \rangle$ of the groundstate $|0\rangle$ for $\vec\theta=(0,\pi)$. The reference site `0' is visually highlighted. Axes carry units $a=|\vec v_1|$, $h=\sqrt{3/4}\,|\vec v_1|$. (b) Density $\langle\hat n_r \rangle$ of the symmetry-broken crystal state $|S\rangle$ (see main text). (c) Fluctuations $\langle\hat n_r^2 \rangle - \langle n_r \rangle^2$ (blue circles), and condensate wavefunction $\vec v_0$ of $|S\rangle$ (red arrows). (d) Scaling of the condensate fraction $\lambda_0/N$ against $N^{-1}$, for superposition states $|S\rangle$ and pinned states with pinning centers of strength $V_p$ on crystal sites.}
\label{fig:spectral-flow}
\end{center}
\end{figure*}

To study the behaviour of the model at densities $\nu>1/3$, we calculate its spectrum and eigenstates using exact numerical diagonalization, focusing on the hardcore limit $u\gg 1$. 
We consider a finite size system consisting 
of $N_b$ bosons in $L_x\times L_y$ rectangular unit cells of the structure shown in Fig.~\ref{fig:EffectiveModel}(a). 
We consider periodic boundary conditions allowing arbitrary flux $(\theta_x,\theta_y)$ passing through the two cycles of the ensuing torus.

We first ask at which densities there can appear phases that are stable to phase separation. This requires positive compressibility, such that $\Delta^2E(N) \equiv E(N+1)+E(N-1)-2E(N)>0$.
The dependency $E(\nu)$ for $\nu_c<\nu<1$ is shown in 
Fig.~\ref{fig:GSEnergy}(a) and $\Delta^2E$ in Fig.~\ref{fig:GSEnergy}(b). 
Besides the crystal described above at $\nu_c=1/3$, three further densities $\nu=1/2$, $\nu=2/3$, 
and $\nu=1$ emerge as natural candidates for stable phases, signalled by a sharp peak in $\Delta^2E$.
Our results for the energy per particle in Fig.~\ref{fig:GSEnergy}(a) indicate that the state of intermediate density $\nu=2/3$ is merely a local minimum and is unstable to phase separation into regions of density $\nu=1/2$ and $1$ at long timescales, as demonstrated in the inset by a Maxwell construction for the energy of a phase 
separated system composed of the adjacent stable phases. 
The physics of the $\nu=1$ state is readily identified as that of a Mott insulator. %

The presence of the additional stable state at $\nu=1/2$ is an intriguing feature of the projected model (\ref{fig:EffectiveModel}). 
Our exact diagonalization studies at this filling, for systems up to $16$ particles, reveal a considerable dependency on the boundary conditions $\vec \theta$.
The eigenvalues disperse, and bands cross, so there is no protected groundstate manifold. %
Nonetheless, at the energy minimum, 
[occurring at $\vec\theta=(0,\pi),\,(\pi,\pi)$, and $(0,0)$ for $4\times 2$, $6\times 2$, and $4\times 2$ lattices, respectively] 
the lowest four eigenvalues %
are separated from higher excited states by an amount larger than their respective splitting.
The correlation function of the lowest eigenstate at this point [Fig.~\ref{fig:spectral-flow}(a)] reveals a triangular crystalline order with lattice vectors $\{2\vec v_1,  2 \vec v_2\}$ 
that suggests a superlattice cell composed of 4 sites, consistent with the relative isolation of the four low-lying eigenstates.

Taking a superposition of the four low-lying (and translationally invariant) exact eigenstates, one can explicitly construct symmetry-broken states \cite{Moller:2010p828}. 
The picture that emerges from these states is that of a supersolid, with half the particles forming a crystal while the remaining particles Bose condense in a state that fills the channels between the former. With hindsight, the presence of a condensate fraction in the system explains the stiffness of the state to a twist in the periodic boundary conditions $\vec\theta$ observed in the exact spectra.

The density matrix $\rho$ for a simple crystalline state of all $N_b$ bosons would be characterized by $N_b$ eigenvalues of order one. 
For the superposition states, we typically find one eigenvalue, $\lambda_0$, that is larger than one and $N_b/2$ eigenvalues of order one, 
signalling the presence of both superfluid and crystalline components.
Generalizing the construction of symmetry-broken condensate states in Ref.~\cite{Moller:2010p828}, we generate these supersolid states as the superpositions of the four lowest 
eigenstates in the exact spectrum by maximizing $\rho_{(k)}\equiv \sum_{i=0}^{k-1}\lambda_i$, the sum of the first $k=N/2+1$ eigenvalues of $\rho$. 
In Fig.~\ref{fig:spectral-flow}(b), we show the density profile for the symmetry-broken state $|S\rangle$ obtained by optimizing $\rho_{(N/2+1)}$, with obvious crystalline order. 
In this state, only half the density is concentrated on superlattice sites, while the remainder forms a condensed background liquid. On the crystal sites, particle number fluctuations are negligible, and there is a very small amplitude for the eigenvector of the largest density matrix eigenvalue on these sites [Fig.~\ref{fig:spectral-flow}(c)]. This eigenvector has support on the background sites, where simultaneously fluctuations are strong. These features are characteristic of a supersolid \cite{Huber:2010p814}. We analyze $\lambda_0$ for the superposition states $|S\rangle$ constructed above, and alternatively for eigenstates obtained using trapping potentials $V_p<0$ on the superlattice sites. Indeed, our numerics indicate the condensate fraction extrapolates to a nonzero value $\simeq 0.06$ in the thermodynamic limit [Fig.~\ref{fig:spectral-flow}(d)].
Unlike the vortex lattices at high density, 
we find that the condensed fraction of this supersolid does not break time-reversal symmetry. 
Experimentally, such crystalline order could be observed clearly in {\it in situ} images of the system; expansion images should reveal evidence for a nonzero condensate fraction, and the density ordering will appear in the noise correlations \cite{Altman:2004p2220}.

In summary, we have proposed the dice-lattice model as an attractive experimental system to explore the many-body physics
of time-reversal symmetric flatbands. We have derived the effective Hamiltonian for interacting particles in such a band.
Our results give the first insights into its rich phase diagram for bosons, driven entirely by interaction-assisted hopping processes
which emerge as a generic feature of projected flat-band Hamiltonians.

\begin{acknowledgments}
We acknowledge support from Trinity Hall Cambridge (G.M.) and EPSRC under EP/F032773/1 (N.R.C.).
\end{acknowledgments}

\bibliography{dice-lattice}

\begin{thebibliography}{36}
\bibitem [{Not()}]{NoteLocalization}%
  \BibitemOpen
  \href@noop {} {}\bibinfo {note} {Note that this property is unlike flat bands
  with non-zero Chern number, in which states cannot be fully
  localized.}\BibitemShut {Stop}%
\end{thebibliography}

\end{document}